\documentclass[twoside,11pt]{article}

\usepackage{graphicx,subfigure}
\usepackage[top=1in,bottom=1in,left=1in,right=1in]{geometry}
\usepackage[sort&compress]{natbib} 
\usepackage{amsfonts, amsmath, amssymb, amsthm, constants, bbm,mathrsfs}
\usepackage{algorithm,algpseudocode}
\usepackage{MnSymbol}
\usepackage{multirow, booktabs}
\usepackage{multirow}
\usepackage{array}
\newcolumntype{L}[1]{>{\raggedright}m{#1}}
\newcolumntype{C}[1]{>{\centering}m{#1}}
\newcolumntype{R}[1]{>{\raggedleft}m{#1}}

\usepackage{titlesec}
\titleformat*{\paragraph}{\large\it}
\usepackage{color}
\definecolor{darkred}{RGB}{100,0,0}
\definecolor{darkgreen}{RGB}{0,100,0}
\definecolor{darkblue}{RGB}{0,0,150}

\usepackage{hyperref}
\hypersetup{colorlinks=true, linkcolor=darkred, citecolor=darkgreen, urlcolor=darkblue}
\usepackage{url}

\def\cond{\,\vert\,}

\newtheorem{thm}{Theorem}
\newtheorem{prp}{Proposition}
\newtheorem*{prp*}{Proposition}

\theoremstyle{remark}

\newtheorem{rem}{Remark}

\def\beq{\begin{equation}} % \setcounter{equation}{1}}
\def\eeq{\end{equation}}
\def\beqn{\begin{eqnarray*}}
\def\eeqn{\end{eqnarray*}}
\def\Bitem{\begin{itemize}\setlength{\itemsep}{.2in}}
\def\bitem{\begin{itemize}\setlength{\itemsep}{.05in}}
\def\eitem{\end{itemize}}
\def\Benum{\begin{enumerate}\setlength{\itemsep}{.2in}}
\def\benum{\begin{enumerate}\setlength{\itemsep}{.05in}}
\def\eenum{\end{enumerate}}
\def\bmult{\begin{multline*}}
\def\emult{\end{multline*}}
\def\bcenter{\begin{center}}
\def\ecenter{\end{center}}
\def\bframe{\begin{frame}}
\def\eframe{\end{frame}}

\newcommand{\thmref}[1]{Theorem~\ref{thm:#1}}

\newcommand{\secref}[1]{Section~\ref{sec:#1}}
\newcommand{\figref}[1]{Figure~\ref{fig:#1}}
\newcommand{\tabref}[1]{Table~\ref{tab:#1}}
\DeclareMathOperator*{\argmax}{arg\, max}
\DeclareMathOperator*{\argmin}{arg\, min}
\def\bP{\mathbf{P}}

\newcommand{\bmu}{{\boldsymbol\mu}}

\def\bbI{\mathbb{I}}

\def\bbP{\mathbb{P}}

\def\bbR{\mathbb{R}}

\newcommand{\E}{\operatorname{\mathbb{E}}}

\def\1{\mathbbm{1}}
\newcommand{\IND}[1]{\bbI\{ #1 \}}

\definecolor{purple}{rgb}{0.4,.1,.9}

\newcommand\blfootnote[1]{%
  \begingroup
  \renewcommand\thefootnote{}\footnote{#1}%
  \addtocounter{footnote}{-1}%
  \endgroup
}

\def\bpi{\boldsymbol\pi}

\begin{document}
\thispagestyle{empty}

\title{Semiparametric Estimation of Symmetric Mixture Models with Monotone and Log-Concave Densities}
\author{Xiao Pu \and Ery Arias-Castro}
\date{}
\maketitle

\blfootnote{Both authors are with the Department of Mathematics, University of California, San Diego, USA.  Contact \href{http://www.math.ucsd.edu/~xipu/}{Xiao (Victor) Pu} or \href{http://math.ucsd.edu/~eariasca}{Ery Arias-Castro}.
We are grateful to Kaspar Rufibach and Lutz D\"umbgen for their feedback on the adaptation of their work in \secref{npmle}.  We are also grateful to G\"unther Walther for answering questions of identifiability and for calling our attention to the work of \cite{balabdaoui2014location}.
This work was partially supported by a grant from the US Office of Naval Research (N00014-13-1-0257) and a grant from the US National Science Foundation (DMS 1223137).
}  

\vspace{-0.3in}

\begin{abstract}
In this article, we revisit the problem of fitting a mixture model under the assumption that the mixture components are symmetric and log-concave. To this end, we first study the nonparametric maximum likelihood estimation (MLE) of a monotone log-concave probability density. 
To fit the mixture model, we propose a semiparametric EM (SEM) algorithm, which can be adapted to other semiparametric mixture models. In our numerical experiments, we compare our algorithm to that of \cite{balabdaoui2014inference} and other mixture models both on simulated and real-world datasets.

\end{abstract}

\section{Introduction} \label{sec:intro}
Mixture models are a staple of statistical analysis.  In this paper, we concern ourselves with semi-parametric mixture models under the hypotheses of symmetry and log-concavity.

Consider the following mixture model (in dimension 1)
$g(x) = \sum_{j = 1}^k \pi_j f(x - \mu_j)$, where, as usual, $\pi_j \ge 0$ and $\sum_j \pi_j = 1$, $\mu_j \in \bbR$, and $f$ is a density on the real line.  Let $\bpi = (\pi_1, \dots, \pi_k)$ and $\bmu = (\mu_1, \dots, \mu_k)$.
\cite{bordes2006semiparametric} and \cite{hunter2007inference} examine the identifiability of such a model under the assumption that $f$ is symmetric.  Their work shows that the model is identifiable when $k = 2$ (up to labeling) as long as $\pi_1 \neq 1/2$.  They also consider the case of $k = 3$ components and show that identifiability holds except for sets of zero Lebesgue measure.  
\cite{balabdaoui2014location} assume $f$ is a  P\'olya frequency function of infinite order and show that the model is identifiable (as always, modulo relabeling), regardless of whether $f$ is symmetric or not (in the latter case, $f$ is assumed to have zero mean).  

Regardless of identifiability, fitting mixture models to data is common practice in statistics, often as an exploration procedure to uncover interesting features of the underlying distribution (e.g., clusters).
In terms of methods for fitting such models, \cite{bordes2006semiparametric} use the so-called minimum contrast method to estimate $\bpi$ and $\bmu$, and use a kernel density estimation (KDE) approach which involves a model selection procedure to choose the tuning parameter. \cite{hunter2007inference} employ a generalized Hodges-Lehmann estimator to estimate $\bmu$ and achieve a better rate of convergence. However, their estimator for $f$ is not guaranteed to be a density.  \cite{bordes2007stochastic} propose a stochastic EM-like estimation algorithm which does not possess the monotone property of a genuine EM algorithm. 
\cite{butucea2014semiparametric} propose $\sqrt{n}$-consistent M-estimators based on a Fourier approach.

\cite{balabdaoui2014inference} consider the same model with $f$ symmetric and log-concave.  Their method for fitting the model consists in adopting the estimators for $\bpi$ and $\bmu$ from \citep{hunter2007inference} and then estimating the density $f$ via maximum likelihood.
\cite{chang2007clustering} study a more general model, $g(x) = \sum_{j = 1}^k \pi_j f_j(x)$, where each $f_j$ is assumed to be log-concave.  They provide an EM-type algorithm for fitting such a model --- however, they do not prove that their algorithm increases the likelihood with every iteration.  \cite{hu2016maximum} study the theoretical properties of the maximum likelihood estimator (MLE) for this same model, although the model is not identifiable as argued in\citep{walther2002detecting}.

In the present paper, we consider fitting a mixture model of the form 
\beq
\label{our-model}
g(x) = \sum_{j = 1}^k \pi_j f_j(x - \mu_j),  \quad \sum_{j=1}^k \pi_j = 1, \quad x \in \bbR,
\eeq
where each $f_j$ is assumed symmetric and log-concave.  
%We propose a direct maximum likelihood approach and design a genuine EM algorithm with the usual monotonicity property.  
%Our approach is based on the maximum likelihood estimator for a monotone log-concave density.  
A density $f$ that is symmetric and log-concave is defined by $h(x) = 2 f(x) 1_{\{x \ge 0\}}$, which is a decreasing log-concave density on $\bbR_+$.  
We make no claims as regards identifiability of this model.  Our goal is to simply fit such a model to data.

In \secref{npmle} we start by examining the maximum likelihood estimation of a symmetric and log-concave density, relying heavily on the work of \cite{rufibach2006log} and \cite{dumbgen2009maximum}.
In \secref{SEM} we propose a genuine EM algorithm for fitting the mixture model \eqref{our-model}.  The algorithm includes a step where the monotone and log-concave MLE for $h_j$ is computed.  To do so we apply the method\footnote{ The method is based on an active set implementation and is available in the R package {\sf logcondens.mode}.} of \cite{doss2016mode} designed for computing the log-concave MLE with a fixed mode --- the mode is of course set to 0 in our case.  We note that \cite{balabdaoui2014inference} use the same routine in the numerical implementation of their method.
In \secref{cluster} we apply our model to clustering problems and compare our approach with that of \citep{chang2007clustering} and that of \citep{balabdaoui2014inference}, as well as a Gaussian mixture model, on both synthetic and real-world datasets.

\section{On the maximum likelihood estimation of a monotone and log-concave density}\label{sec:npmle}
This section is concerned with the maximum likelihood estimation of a monotone log-concave density on $\bbR_+$.
Maximum likelihood estimation of a monotone density was first studied by \cite{grenander1956theory}, while the maximum likelihood estimation of a log-concave density has garnered attention only more recently \citep{balabdaoui2004nonparametric, doss2016global, balabdaoui2009limit, dumbgen2009maximum, walther2002detecting}.
Our exposition and results below derive from a straightforward adaptation of the thesis work of \cite{rufibach2006log} on the maximum likelihood of a log-concave density, without the additional constraint of monotonicity, published in the form of a research article \citep{dumbgen2009maximum}.  
We do not provide proofs but rather refer the reader to that work for all the technical details.

Let $f$ denote a decreasing and log-concave density on $\bbR_+$.  We let $F$ denote the distribution function corresponding to the density $f$ and define 
\beq
\psi(x) = \log f(x).
\eeq
Requiring that $f$ be monotone and log-concave is equivalent to requiring that $\psi$ is monotone and concave.
Based on the sample, which we assume ordered ($x_1 < x_2 < \cdots < x_n$), the negative log-likelihood at $f$ is given by 
\beq
\label{negll}
-\sum_{i=1}^n \log f(x_i) = -n \sum_{i=1}^n \psi(x_i).
\eeq
In order to relax the constraint of $f$ being a probability density we follow the technique used by \cite{rufibach2006log} and add a Lagrange term to \eqref{negll}, leading to the functional
\beq\label{obj0}
\Lambda_n(\psi) = -\sum_{i=1}^n \psi(x_i) + n \int_0^\infty \exp \psi(x) {\rm d}x.
\eeq
The MLE of $f$ is $\hat f_n = \exp \hat \psi_n$, where $\hat\psi_n$ is the minimizer of $\Lambda$ over class of functions on $[0, \infty)$ that are non-increasing and concave, that is
\beq
\hat\psi_n := \argmin_{\psi \in \mathcal{MC}}\, \Lambda_n (\psi),
\eeq
where\footnote{ Following the definition in \cite{rockafellar2015convex}, a concave function $f$ is said to be proper if $f(x) > -\infty$ for at least one $x$ and $f(x) < +\infty$ for every $x$.  A closed function is a function that maps closed sets to closed sets.}
\beq
\mathcal{MC} := \big\{\psi: [0, \infty) \to [-\infty, \infty)\  | \  \psi \text{ is non-increasing, concave, proper, and closed} \big\}.
\eeq

The following results from an adaptation of Theorem~2.1 in \citep{dumbgen2009maximum}.  
\begin{thm}[Existence, uniqueness, and shape] \label{thm:exist}
The MLE $\hat\psi_n$ exists and is unique. It is linear between sample points and continuous on $[0, x_n]$, with $\hat\psi_n(x) = \hat\psi_n(x_1)$ for $x \in [0,x_1]$ and $\hat\psi_n(x) = -\infty$ for $x > x_n$. 
\end{thm}

The following results from an adaptation of Theorem~2.2 in \citep{dumbgen2009maximum}.  
\begin{thm}[Characterization] \label{thm:property}
Let $\psi$ be a non-increasing and concave function such that $\{x: \psi(x) > -\infty\} = [0, x_n]$.
Then, $\psi = \hat\psi_n$ if and only if 
\beq
\frac1n \sum_{i=1}^n \Delta(x_i) \le \int_0^\infty \Delta(x) \exp \psi(x) {\rm d}x
\eeq
for any $\Delta: [0, \infty) \to \bbR$ such that $\psi + \lambda \Delta$ is non-increasing and concave for some $\lambda > 0$.
\end{thm}

For $I \subset \bbR$ an interval, $\beta \in [1,2]$, and $L > 0$, let $\mathcal{H}^{\beta, L} (I)$ be the H\"older class of real-valued functions $g$ on $I$ satisfying $|g(y) - g(x)| < L|y-x|$ if $\beta = 1$ and $|g'(y) - g'(x)| \leq L|y-x|^{\beta -1}$ if $\beta \in (1,2]$, for all $x,y \in I$. 
The following results from an adaptation of Theorem~4.1 in \citep{dumbgen2009maximum}.  
\begin{thm}[Uniform consistency] \label{thm:consistency}
Assume that $f \in \mathcal{H}^{\beta, L} (I)$ for some exponent $\beta \in [1,2]$, some constant $L > 0$, and a compact interval $I \subset \{f>0\}$. Then, 
\beq
\max_{t \in I} \big|\hat f_n(t) - f(t)\big| = O_\bbP(\log n/n)^{\beta/(2\beta + 1)}.
\eeq
\end{thm}

As pointed out by \cite{dumbgen2009maximum}, this is the minimax rate for densities in that smoothness class, as shown by \cite{khas1979lower}, so that, when the density is log-concave and H\"older-$\beta$ (with $\beta \in [1,2]$) in some interval, the log-concave MLE adapts to the proper smoothness in that interval.  We believe the same holds under the additional constraint of monotonicity.

\section{A semiparametric EM algorithm} \label{sec:SEM}
We now consider fitting the semiparametric mixture model \eqref{our-model}.  Model \eqref{our-model} is defined by $\phi := (\bmu; \bpi; \boldsymbol{f})$, where $\bmu = (\mu_1, \dots, \mu_k) \in \bbR^k$, $\bpi = (\pi_1, \dots, \pi_k)$ is an element of the simplex in $\bbR^k$, and $\boldsymbol f = (f_1,\dots, f_k)$ with each $f_j$ being a symmetric and log-concave density on $\bbR$.  
Under $\phi$, the mixture model is given by 
\beq
g_\phi(x) = \sum_{j=1}^k \pi_j f_j(x - \mu_j).
\eeq
Given a sample $\boldsymbol{x} = (x_1, \dots, x_n)$, the  log-likelihood under parameter $\phi$ is given by
\beq
\label{likelihood}
L(\phi) = \sum_{i=1}^n \log g_\phi(x_i).
\eeq
(As is customary, we leave the dependency in $\boldsymbol x$ implicit throughout.)
%For an observation $x$, let $y = (x, z)$ where $z \in \{1, \dots, k\}$ identifies the component $x$ was generated from.  In a clustering setting, the label $z$ is of course missing.  
As with other mixture models, maximizing $L(\phi)$ directly is difficult and we resort to an EM-type approach.  Define the indicator variables $\boldsymbol{z} := (z_1, \dots, z_n)$, where $z_i = j$ when $x_i$ was sampled from the $j$th component.  These allow us to define the complete log-likelihood
\begin{align}
\bar L(\phi; \boldsymbol{z}) = \sum_{i=1}^n\sum_{j=1}^k \IND{z_i = j} \log(f_j(x_i-\mu_j)).
\end{align}
An EM algorithm, after some initialization, alternates between computing the expectation of the complete log-likelihood conditional on $\boldsymbol{x}$ under the current value of the parameters, and maximizing the resulting functional with respect to the parameters.
Thus, in the expectation step, assuming that the current value of the parameters is $\phi^*$, consists in computing
\beq
\E_{\phi^*}[\bar L(\phi; \boldsymbol{z}) \mid \boldsymbol{x}],
\eeq
which turns out to be equal to
\beq
\label{bound-left}
Q(\phi, \boldsymbol{w}^*) 
:= \sum_{i=1}^n\sum_{j=1}^k w_{ij}^* \log(\pi_j f_j(x_i-\mu_j)),
\eeq
%\sum_{i=1}^n\sum_{j=1}^k w_{ij}^* \log(\pi_j) + \sum_{i=1}^n\sum_{j=1}^k w_{ij}^* \log(f_j(x_i-\mu_j)), \label{E-step}
where 
\beq
\label{post}
w_{ij}^*  := \bbP_{\phi^*} (z_i = j \cond x_i) = \frac{\pi_j^* f_j^*(x_i-\mu_j^*)}{\sum_{l=1}^k \pi_l^* f_l^*(x_i-\mu_l^*)}.
\eeq
The maximization step consists then in maximizing $Q(\phi, \boldsymbol{w}^*)$ with respect to $\phi$, thus updating the value of the parameters.
This leads to the following semiparametric EM (SEM) algorithm:
\bitem
\item {\bf E-step}: 
\bitem
\item Compute $\boldsymbol{w}$:
\beq
\label{post-t}
w^t_{ij} \gets \bbP_{\phi^t} (z_i = j \cond x_i) = \frac{\pi^t_j f_j^t(x_i-\mu_j^t)}{\sum_{l=1}^k \pi_l^t f_l^t(x_i-\mu_l^t)},
\eeq
for $ i = 1, \dots, n$ and $ j = 1, \dots, k$.
\eitem
\item {\bf M-step}: 
\bitem
\item Update $\bpi$:
\beq
\pi_j^{t+1} \gets \frac{1}{n} \sum_{i=1}^n w_{ij}^t, \quad  j = 1, \dots, k.
\eeq
\item Update $\bmu$:
\beq
\label{update-mu}
\mu_j^{t+1} \gets \argmax_{\mu \in \bbR}\  \sum_{i=1}^n w_{ij}^t \log f_j^t(x_i-\mu),\quad  j = 1, \dots, k.
\eeq
\item Update $\boldsymbol f$:
\beq
\label{update-f}
f_j^{t+1}(x) \gets \tfrac12 h_j^{t+1}(|x|), \quad
h_j^{t+1} \gets \argmax_{h \in \mathcal{MC}} \  \sum_{i=1}^n w_{ij}^t \log h(|x_i-\mu_j^{t+1}|), \quad  j = 1, \dots, k.
\eeq
\eitem
\eitem
%In \eqref{update-mu}, monotone an\phi^td log-concave density $f_j$ might not be first-derivative continuous, therefore  we numerically update $\mu_j$ through the function {\bf hjk} in a derivative-free optimization R package {\bf dfoptim}. 
As a function of $\mu$, the function appearing in \eqref{update-mu} is concave due to the fact that $\log f_j^t$ is concave by construction.  In particular, the Golden Section Search can be applied to find $\mu$. 
In \eqref{update-f}, the optimization is over $h$ being a monotone and log-concave density on $[0, \infty)$.  The solution corresponds to the weighted MLE based on data $(|x_1 - \mu_j^{t+1}|, \dots, |x_n -\mu_j^{t+1}|)$ and weights $(w_{1j (t)}, \dots, w_{nj (t)})$.  
in our implementation we apply the function {\sf activeSetLogCon.mode} in the R package {\sf logcondens.mode} with mode chosen to be 0.

Our SEM algorithm has the desirable monotonicity property of a true EM algorithm \citep{dempster1977maximum,wu1983convergence}.

\begin{prp}[Monotonicity property]\label{prp:monotone}
We $L(\phi^t) \leq L(\phi^{t+1})$ for all $t \ge 0$.
\end{prp}

\begin{proof}
In the algorithm, armed with $\phi^t$, we compute the weights $\boldsymbol{w}^t$ in the E-step and in the M-step we obtain $\phi^{t+1}$ by maximizing $Q(\phi, \boldsymbol{w}^t)$ over $\phi$.  
%(We do the latter sequentially, first over $\bpi$, then over $\bmu$, and finally over $\boldsymbol f$.)  
In particular,
\beq\label{Q-ineq}
Q(\phi^{t+1}, \boldsymbol{w}^t) \ge Q(\phi^t, \boldsymbol{w}^t).
\eeq

The key in what follows is Jensen's inequality, which implies that for a set of parameters $\phi = (\bmu; \bpi; \boldsymbol h)$ and non-negative weights $\boldsymbol{w}^* = (w^*_{ij})$ such that $\sum_j w^*_{ij} = 1$ for all $i$, 
\begin{align}
\label{bound}
L(\phi) 
&= \sum_{i=1}^n \log \bigg(\sum_{j=1}^k \pi_j f_j(x_i - \mu_j)\bigg) \\
&= \sum_{i=1}^n \log \bigg(\sum_{j=1}^k  w^*_{ij}\frac{\pi_j f_j(x_i-\mu_j)}{w^*_{ij}}\bigg) \nonumber\\
&\ge \sum_{i=1}^n \sum_{j=1}^k  w^*_{ij} \log \bigg(\frac{\pi_j f_j(x_i-\mu_j)}{w^*_{ij}}\bigg) \nonumber\\
&= Q(\phi, \boldsymbol{w}^*) - C(\boldsymbol{w}^*),
\end{align}
where $C(\boldsymbol w) :=  \sum_{i=1}^n\sum_{j=1}^k w_{ij} \log w_{ij}  + n \log2$.  The inequality is in fact an equality if the weights $\boldsymbol{w}^*$ are the weights associated with $\phi$ as specified in \eqref{post}.
In particular,
\beq
L(\phi^{t+1}) 
\ge Q(\phi^{t+1}, \boldsymbol{w}^t) - C(\boldsymbol{w}^t),
\eeq
while
\beq
L(\phi^t) 
= Q(\phi^t, \boldsymbol{w}^t) - C(\boldsymbol{w}^t).
\eeq

With this, together with \eqref{Q-ineq}, we have
\begin{align*}
L(\phi^{t+1}) 
&\ge Q(\phi^{t+1}, \boldsymbol{w}^t) - C(\boldsymbol{w}^t) \\
&\ge Q(\phi^t, \boldsymbol{w}^t) - C(\boldsymbol{w}^t) 
= L(\phi^t). \qedhere
\end{align*}

\end{proof}

\begin{rem}[Initialization] 
In practice, we initialize $\boldsymbol{w}^0$ and $\boldsymbol h^0$ at the values computed by fitting a Gaussian mixture model (using an EM algorithm) and start with M-step first.
\end{rem}

\section{Numerical experiments} \label{sec:cluster}
We present in this section the result of some numerical experiments.
%We assume that the data can be clustered into $k$ groups, fit the $k$-component mixture \eqref{our-model} as described in \secref{SEM} obtaining $\hat\phi$, and assign a label to an observation $x_i$ according to rule
%\beq
%\argmax_{j = 1,\dots,k}\  \bbP (z_i = j | x_i, \hat\phi) = \argmax_{j = 1,\dots,k}\   \frac{\hat\pi_j \hat h_j(|x_i-\hat\mu_j|)}{\sum_{j=1}^k \hat\pi_l \hat h_l(|x_i-\hat\mu_l|)}.
%\eeq 
We apply our SEM algorithm both on simulated and real data. 
In \secref{syn-data} we simulate data from the Gaussian and Laplace mixture models used in \citep{balabdaoui2014inference}, and in \secref{real-data} we apply the SEM algorithm to the well-known Old Faithful Geyser dataset, as done in \citep{balabdaoui2014inference}. 

\subsection{Synthetic datasets}\label{sec:syn-data}
\begin{figure}
\centering\small
\subfigure[$\pi_1^t$ as a function of the iteration $t$ (red) and true value (green)]{
	\includegraphics[scale=0.2]{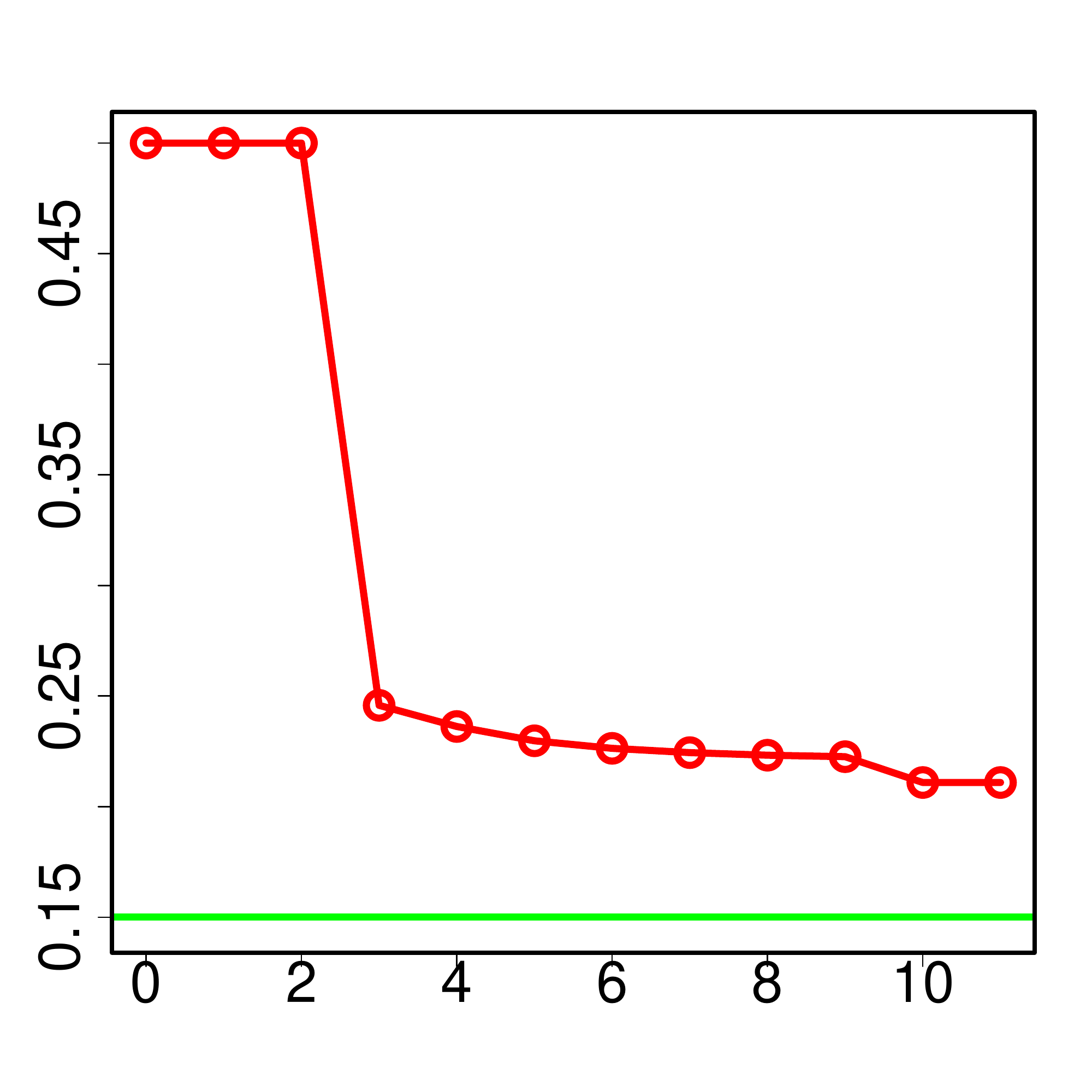}}\hspace{1.5em}
\subfigure[$\mu_1^t$ as a function of the iteration $t$ (red) and true value (green)]{
	\includegraphics[scale=0.2]{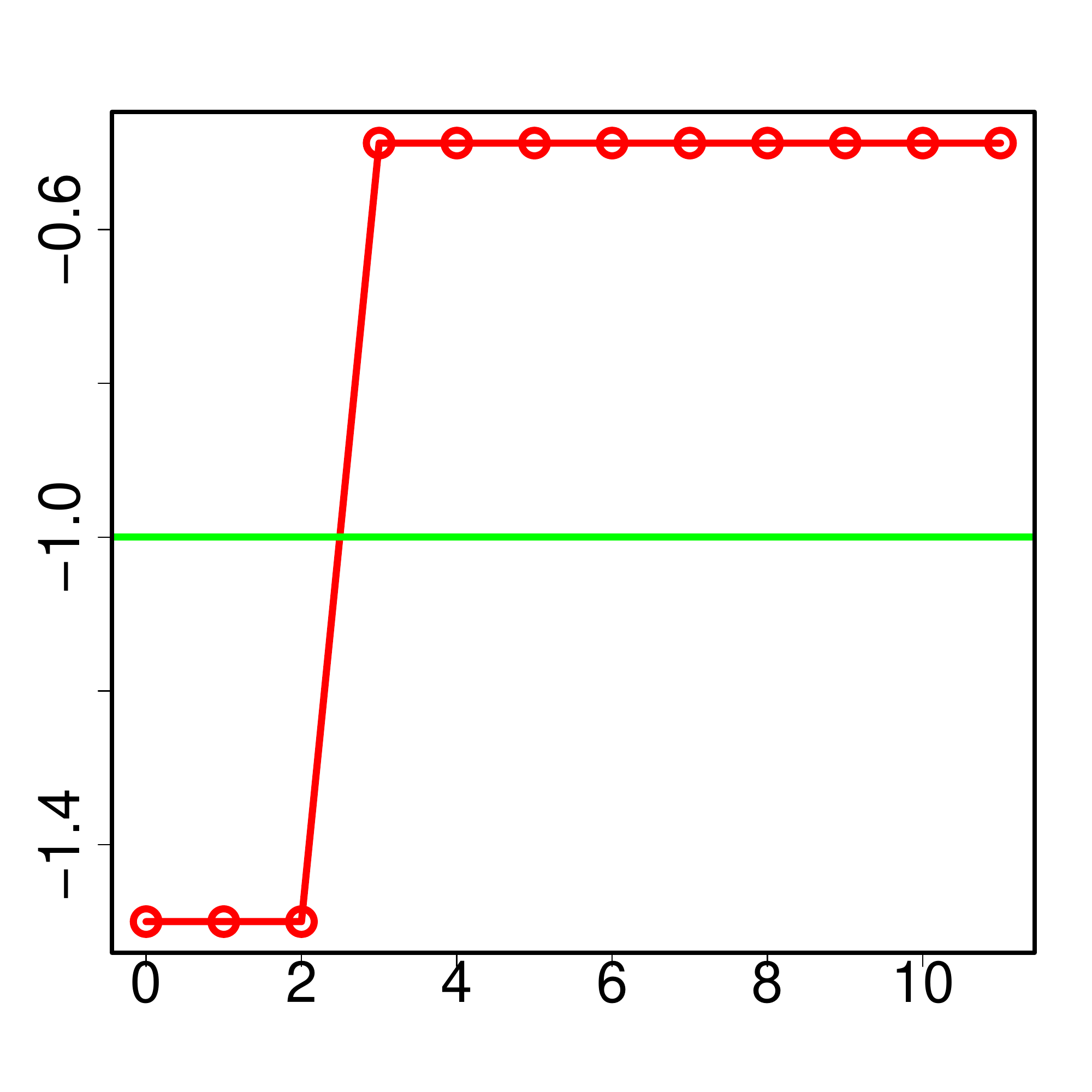}}\hspace{1.5em}
\subfigure[$\mu_2^t$ as a function of the iteration $t$ (red) and true value (green)]{
	\includegraphics[scale=0.2]{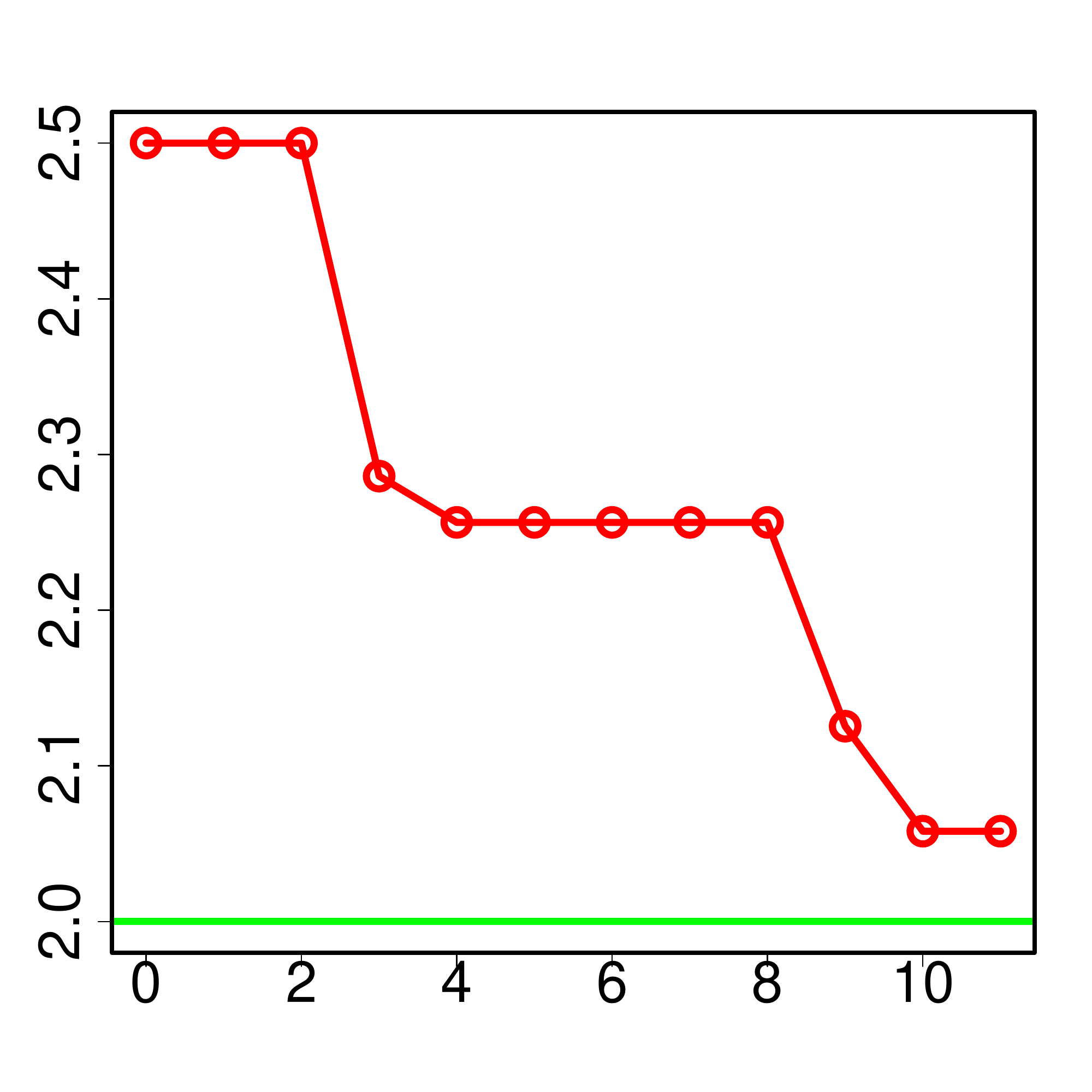}}
	
\subfigure[$L(\phi^t)$ as a function of the iteration $t$]{
	\includegraphics[scale=0.2]{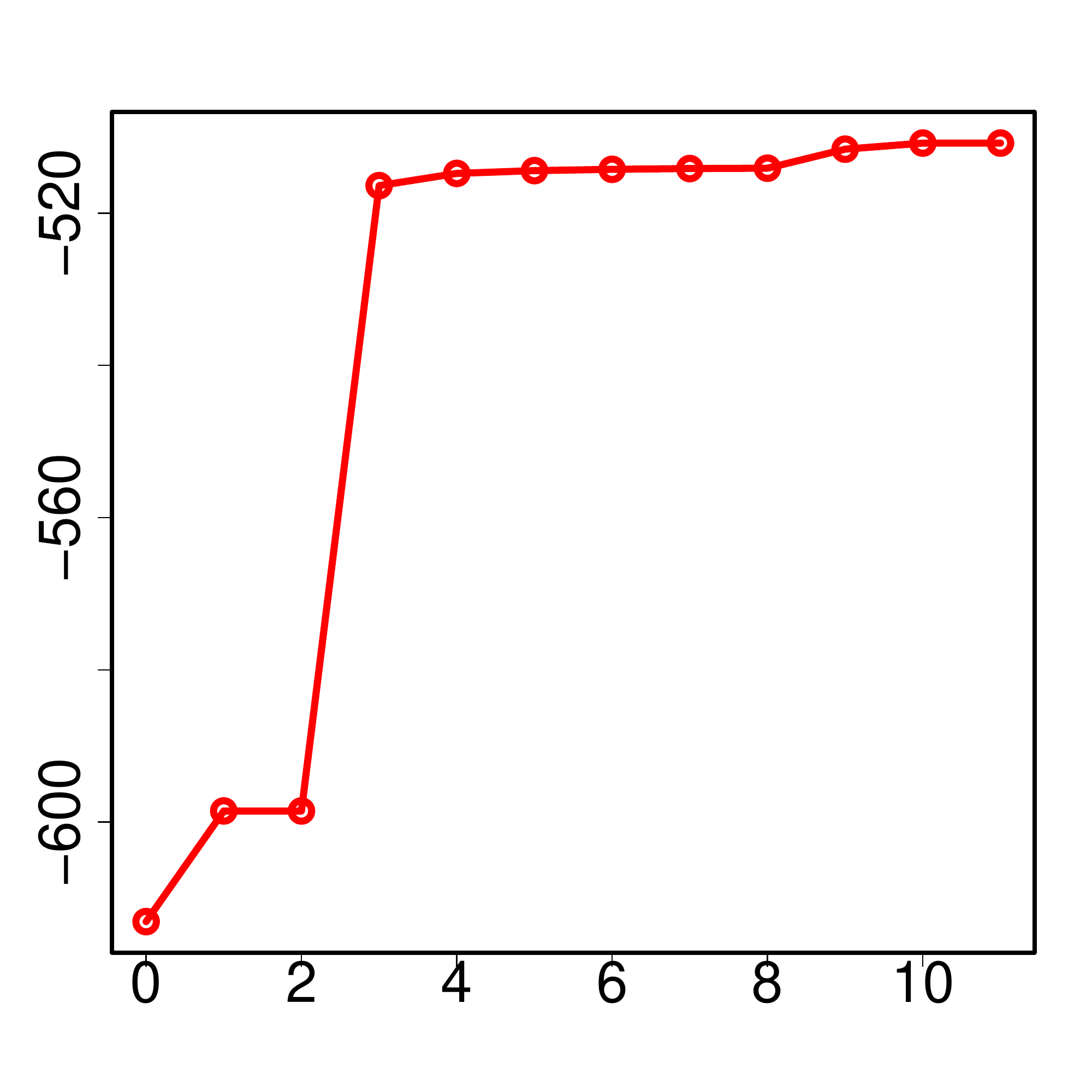}}\hspace{1.5em}
\subfigure[ estimated (red, dashed) and true (green, solid) component densities]{
	\includegraphics[scale=0.2]{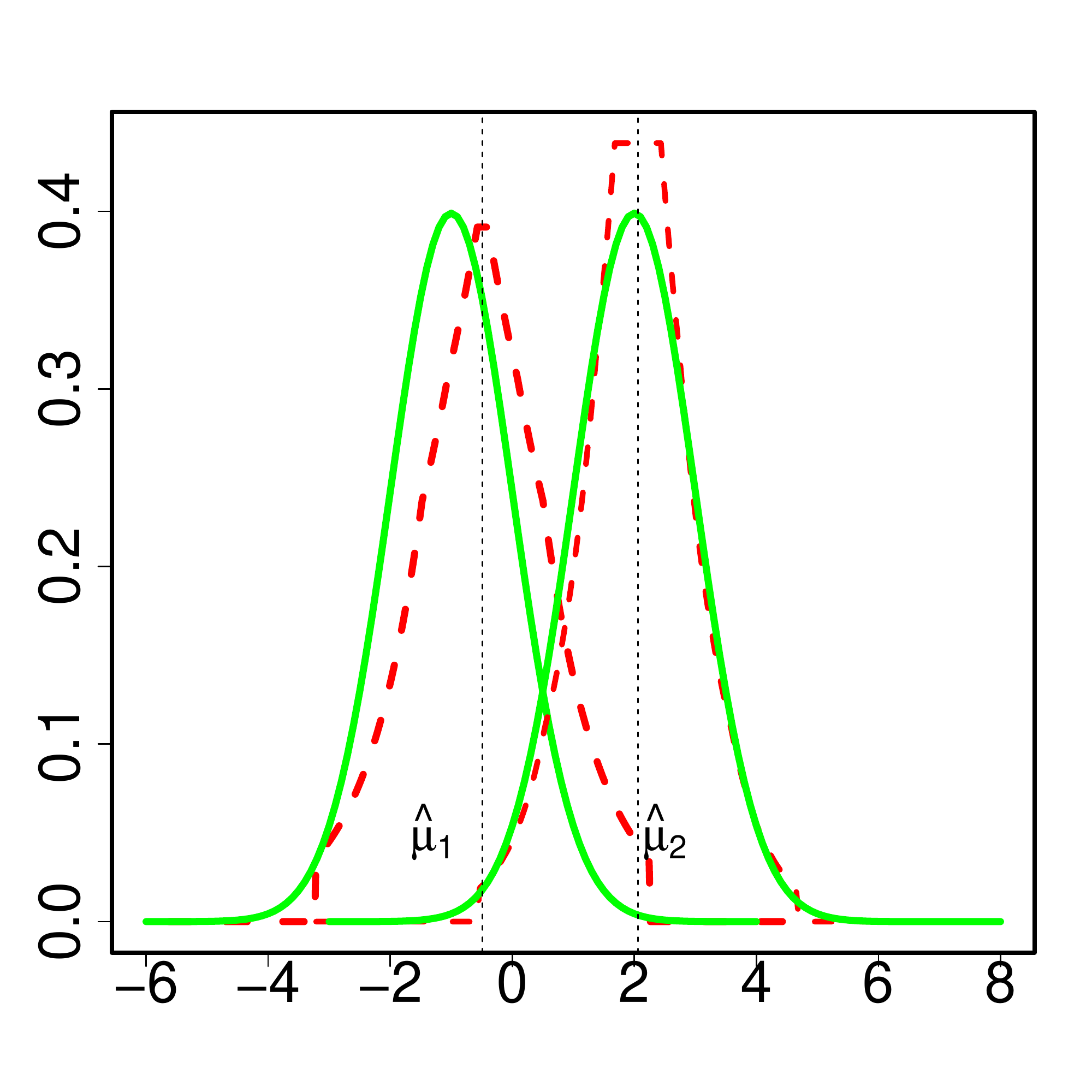}}\hspace{1.5em}
    \subfigure[estimated (red, dashed) and true (green, solid) mixture density]{
	\includegraphics[scale=0.2]{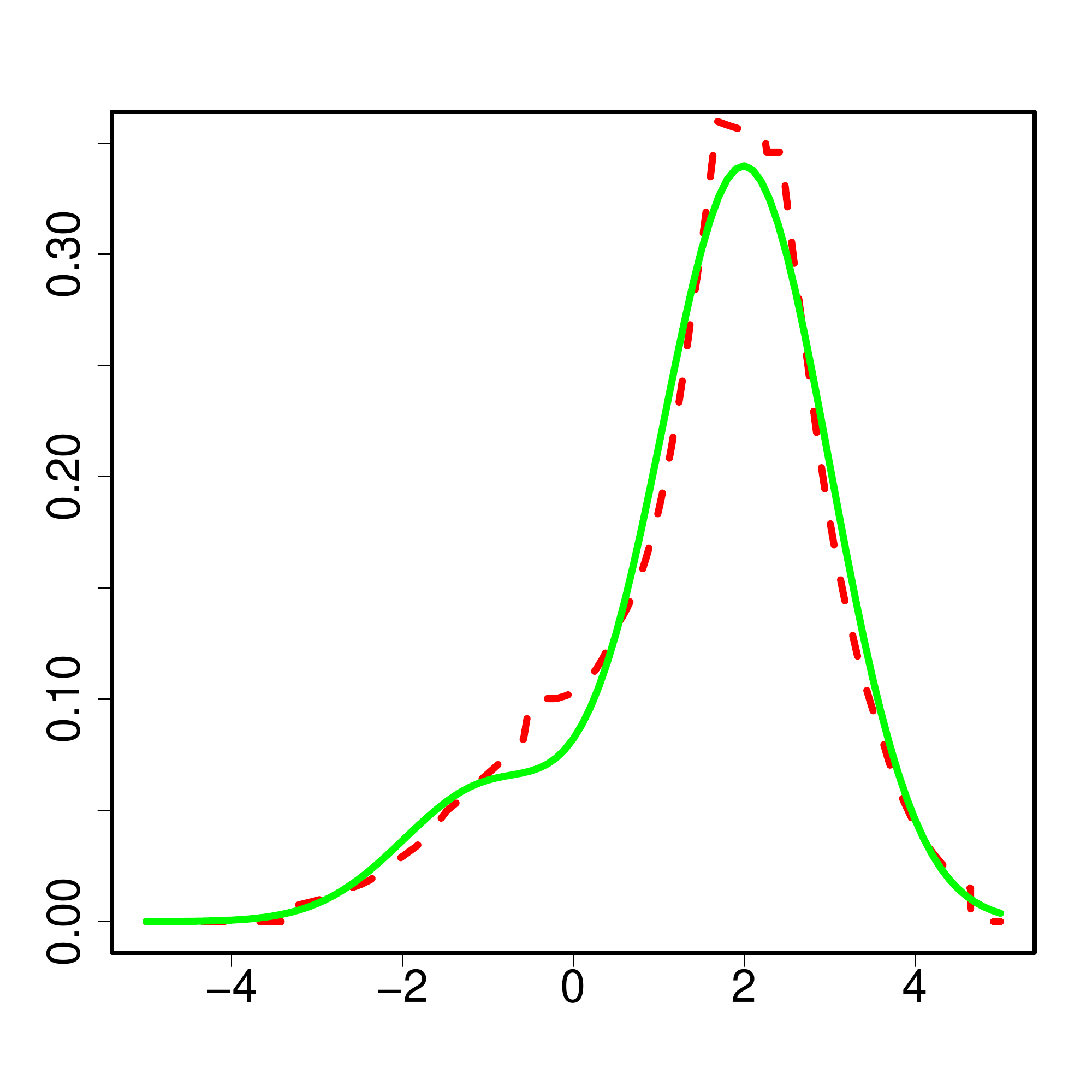}}
\caption{\small SEM for the Gaussian mixture with parameters $k = 2, \pi_1 = 0.15, \mu_1 = -1, \mu_2 = 2$ based on a sample of size $n = 300$.}
\label{fig:gaussian300}
\end{figure}

As a first example, we use a two-component Gaussian mixture to empirically check the convergence of our SEM algorithm. We sample $n = 300$ observations from the Gaussian mixture $0.15 \ \mathcal{N}(-1,1) + 0.85\  \mathcal{N} (2,1)$ and apply the SEM algorithm to these two datasets respectively. This seems to be the most difficult situation considered in \citep{bordes2006semiparametric}.
Panels (a), (b), (c), and (d) of \figref{gaussian300} show that SEM stabilizes after about 8 iterations for the three Euclidean parameters and the observed data likelihood.  
As expected, the achieved maximum data likelihood is monotonically increasing as a function of the number of iterations.  
Panels (e) and (f) show the final MLE for $f_1, f_2, g$ and compare that with the truth.  The MLE for the symmetric log-concave densities are piecewise exponential, which is consistent with what is described in \thmref{exist}. 

We then compare the performance of our algorithm (SEM) for the problem of clustering with the methods proposed in \citep{chang2007clustering}  and \citep{balabdaoui2014inference}. \citeauthor{chang2007clustering} do not assume symmetry while \citeauthor{balabdaoui2014inference} assume that the component densities are identical.  We denote these two methods by LCM and SLC respectively. 
We compare SEM, LCM, SLC, and GMM.  The latter serves as benchmark when the underlying model is a Gaussian mixture.
We compare these methods on two Gaussian mixture models, two Laplace mixture models, and one Gaussian-Laplace mixture model, as described below:
\bitem
\item Model 1: $ 0.2\, \mathcal{N}(0,1) + 0.8\, \mathcal{N}(1,1)$; 
\item Model 2: $0.2\, \mathcal{N}(0,1) + 0.8\, \mathcal{N}(2,2)$;
\item Model 3: $0.2\, \mathcal{L}(0,1) + 0.8\, \mathcal{L}(1,1)$;
\item Model 4: $0.2\, \mathcal{L}(0,1) + 0.4\, \mathcal{L}(1.5,1) + 0.4\, \mathcal{L}(-1.5,1)$;
\item Model 5: $0.2\, \mathcal{N}(0,1) + 0.2\, \mathcal{N}(1.5,1) + 0.2\, \mathcal{N}(-1.5,1) + 0.2\, \mathcal{L}(3,1) + 0.2\, \mathcal{L}(-3,1)$.
\eitem
The sample size is $n = 500$ for Models 1-4, while $n = 1000$ for Model $5$.  Each setting is repeated 1000 times.  We examine the quality of the resulting clustering in terms of the achieved data log-likelihood, the misclassification errors when $k = 2$ or Rand Indexes when $k \ge 3$, and the average absolute posterior probability error used by \citep{chang2007clustering} --- all averaged over the 1000 repeats. 
The latter metric investigates how well a mixture clustering algorithm estimates the uncertainty for the membership assignment of each observation on population level. This metric is defined as
\beq
\label{ab-post}
\text{posterior error} := \frac{1}{n} \sum_{i=1}^n |\hat w_{i1} - w_{i1}|,
\eeq
where $\hat w_{i1}$ and  $w_{i1}$ are computed by \eqref{post} with estimators and true parameters respectively. Notice that this metric only applies to clustering with $k = 2$ components. When $k \geq 3$, we define the posterior error using the Frobenius norm,
\beq
\label{post-error-k}
\text{posterior error} : = \min_{\bP} \| \hat {\boldsymbol w} \bP - \boldsymbol{w} \|_F,
\eeq
where the minimum is over $k \times k$ permutation matrices $\bP$, and matrices $\hat {\boldsymbol w}$ and $\boldsymbol{w}$ are computed by \eqref{post} based on the MLE and true parameters, respectively. We report the comparison results in \tabref{3-mixture}. As can be seen from this table, GMM, LCM, and our SEM algorithm clearly outperform SLC in terms of log-likelihood and posterior error. LCM outperforms other methods in terms of misclassification error or Rand index when $k \le3$, but does not perform well when $k = 5$. When the mixture densities are normal, SEM performs as well as GMM, arguably the gold standard in such a situation; when the densities are Laplace, SEM slightly improves the clustering initialized by GMM. Moreover, SEM achieves a significantly higher log-likelihood compared with the other methods when the mixture densities are normal. We also notice that SLC sometimes gives better results in terms of misclassification error, even though the posterior-error is worse.
%(the Posterior-error resulted from SEM is slightly worse than GMM, but the differences are neglectable for both Gaussian mixtures).
 % Normal mixture N(0,1) + N(1,1)
%102.5710000    0.2058538 -739.3913560
%1.487284429 0.001872138 0.508652871
% N(0,1) + N(2,2)
%125.3980000     0.2548721 -1046.3412411
%1.909240895 0.002641028 0.489151038
% Laplace mixture
%111.397000    0.244285 -869.338000
% 1.226336533 0.001970248 0.685572128

\begin{table}[h]
\footnotesize\centering
\begin{tabular}{C{2.2cm}C{2.4cm}C{2.3cm}C{2.2cm}C{2.2cm}C{2.3cm}}
  \multicolumn{1}{C{2.2cm}}{} &
  \multicolumn{1}{C{2.2cm}}{} &
  \multicolumn{1}{C{2.3cm}}{GMM} &
  \multicolumn{1}{C{2.2cm}}{LCM} &
  \multicolumn{1}{C{2.2cm}}{SLC} &
  \multicolumn{1}{C{2.3cm}}{SEM} \\
\toprule
  \multicolumn{1}{C{2.2cm}}{\multirow{3}{*}{Model 1} }&
  \multicolumn{1}{C{2.4cm}}{log-likelihood} &
  \multicolumn{1}{C{2.3cm}}{-743.2 (0.50)} &
  \multicolumn{1}{C{2.2cm}}{--739.4 (0.51)} &
  \multicolumn{1}{C{2.2cm}}{-1104.9 (2.06)} &
  \multicolumn{1}{C{2.3cm}}{\bf{-738.6 (0.50)}} \\
  
  \multicolumn{1}{C{2.2cm}}{} &
  \multicolumn{1}{C{2.2cm}}{mis-class} &
  \multicolumn{1}{C{2.3cm}}{122.7 (1.31)} &
  \multicolumn{1}{C{2.2cm}}{{\bf 102.6 (1.49)}} &
  \multicolumn{1}{C{2.2cm}}{174.2 (1.22)} &
  \multicolumn{1}{C{2.3cm}}{123.7 (1.30)} \\

  \multicolumn{1}{C{2.2cm}}{} &
  \multicolumn{1}{C{2.2cm}}{post-error} &
  \multicolumn{1}{C{2.3cm}}{\bf{0.199 (0.003)}} &
   \multicolumn{1}{C{2.2cm}}{0.206 (0.002)} &
  \multicolumn{1}{C{2.2cm}}{0.317 (0.001)} &
  \multicolumn{1}{C{2.3cm}}{0.202 (0.003)} \\
\midrule
  \multicolumn{1}{C{2.2cm}}{\multirow{3}{*}{Model 2} } &
  \multicolumn{1}{C{2.4cm}}{log-likelihood} &
  \multicolumn{1}{C{2.3cm}}{-1049.7 (0.48)} &
   \multicolumn{1}{C{2.2cm}}{-1046.3 (0.49)} &
  \multicolumn{1}{C{2.2cm}}{-1383.0 (2.43)} &
  \multicolumn{1}{C{2.5cm}}{{\bf-1044.7 (0.48)}} \\
  
  \multicolumn{1}{C{2.2cm}}{} &
  \multicolumn{1}{C{2.2cm}}{mis-class} &
  \multicolumn{1}{C{2.3cm}}{148.1 (1.55)} &
  \multicolumn{1}{C{2.3cm}}{125.4 (1.91)} &
  \multicolumn{1}{C{2.2cm}}{\bf{118.4 (0.70)}} &
  \multicolumn{1}{C{2.3cm}}{150.0 (1.54)} \\
  
  \multicolumn{1}{C{2.2cm}}{} &
  \multicolumn{1}{C{2.2cm}}{post-error} &
  \multicolumn{1}{C{2.3cm}}{{\bf 0.211 (0.004)}} &
  \multicolumn{1}{C{2.2cm}}{0.255 (0.003)} &
  \multicolumn{1}{C{2.2cm}}{0.216 (0.004)} &
  \multicolumn{1}{C{2.3cm}}{0.283 (0.001)} \\
\midrule

  \multicolumn{1}{C{2.2cm}}{\multirow{3}{*}{Model 3} } &
  \multicolumn{1}{C{2.4cm}}{log-likelihood} &
  \multicolumn{1}{C{2.3cm}}{ -876.0 (0.67)} &
  \multicolumn{1}{C{2.3cm}}{{\bf -869.3 (0.69)} } &
  \multicolumn{1}{C{2.2cm}}{-1293.6 (3.94)} &
  \multicolumn{1}{C{2.3cm}}{ -870.6 (0.66) } \\
  
  \multicolumn{1}{C{2.2cm}}{} &
  \multicolumn{1}{C{2.2cm}}{mis-class} &
  \multicolumn{1}{C{2.3cm}}{162.8 (1.12)} &
  \multicolumn{1}{C{2.2cm}}{ {\bf 111.4 (1.23)}} &
  \multicolumn{1}{C{2.2cm}}{153.2 (1.09)} &
  \multicolumn{1}{C{2.3cm}}{159.1 (1.18)} \\
  
  \multicolumn{1}{C{2.2cm}}{} &
  \multicolumn{1}{C{2.2cm}}{post-error} &
  \multicolumn{1}{C{2.3cm}}{0.236 (0.002)} &
   \multicolumn{1}{C{2.2cm}}{0.244 (0.002)} &
  \multicolumn{1}{C{2.2cm}}{0.324 (0.001)} &
  \multicolumn{1}{C{2.3cm}}{{\bf 0.234 (0.002)}} \\
  \midrule

  \multicolumn{1}{C{2.2cm}}{\multirow{3}{*}{Model 4} } &
  \multicolumn{1}{C{2.4cm}}{log-likelihood} &
  \multicolumn{1}{C{2.3cm}}{ -1031.1 (16.8)} &
  \multicolumn{1}{C{2.3cm}}{-1030.0 (16.6) } &
  \multicolumn{1}{C{2.2cm}}{-} &
  \multicolumn{1}{C{2.3cm}}{ {\bf -1025.7 (16.8)}} \\
  
  \multicolumn{1}{C{2.2cm}}{} &
  \multicolumn{1}{C{2.2cm}}{Rand index} &
  \multicolumn{1}{C{2.3cm}}{0.602 (0.097)} &
  \multicolumn{1}{C{2.2cm}}{ {\bf 0.655 (0.118)}} &
  \multicolumn{1}{C{2.2cm}}{-} &
  \multicolumn{1}{C{2.3cm}}{0.607 (0.093)} \\
  
  \multicolumn{1}{C{2.2cm}}{} &
  \multicolumn{1}{C{2.2cm}}{post-error} &
  \multicolumn{1}{C{2.3cm}}{{\bf10.6 (2.65)}} &
   \multicolumn{1}{C{2.2cm}}{13.3 (1.53)} &
  \multicolumn{1}{C{2.2cm}}{-} &
  \multicolumn{1}{C{2.3cm}}{{\bf 10.6 (2.61)}} \\
  \midrule

  \multicolumn{1}{C{2.2cm}}{\multirow{3}{*}{Model 5} } &
  \multicolumn{1}{C{2.4cm}}{log-likelihood} &
  \multicolumn{1}{C{2.3cm}}{ -2281.8 (19.6)} &
  \multicolumn{1}{C{2.3cm}}{-2282.9 (20.4) } &
  \multicolumn{1}{C{2.2cm}}{-} &
  \multicolumn{1}{C{2.3cm}}{ {\bf -2276.6 (19.8)}} \\
  
  \multicolumn{1}{C{2.2cm}}{} &
  \multicolumn{1}{C{2.2cm}}{Rand index} &
  \multicolumn{1}{C{2.3cm}}{0.617 (0.134)} &
  \multicolumn{1}{C{2.2cm}}{ 0.401(0.231)} &
  \multicolumn{1}{C{2.2cm}}{-} &
  \multicolumn{1}{C{2.3cm}}{{\bf 0.619 (0.132)}} \\
  
  \multicolumn{1}{C{2.2cm}}{} &
  \multicolumn{1}{C{2.2cm}}{post-error} &
  \multicolumn{1}{C{2.3cm}}{{\bf 18.2 (3.13)}} &
   \multicolumn{1}{C{2.2cm}}{19.8 (3.25)} &
  \multicolumn{1}{C{2.2cm}}{-} &
  \multicolumn{1}{C{2.3cm}}{18.4 (3.21)} \\
\bottomrule
\end{tabular}
\caption{\small
Comparison of the four different clustering methods in terms of achieved log-likelihood, number of misclassification errors (when $k = 2$) or Rand index (when $k > 2$), and posterior errors.  The reported numbers are averages (and standard deviations) based on 1000 replications.  The sample size is $n = 500$ for Model 1-4 and $n = 1000$ for Model 5.  (SLC was only designed for mixtures with $k=2$ components.)} 
\label{tab:3-mixture}
\end{table}

\subsection{Real dataset} \label{sec:real-data}
In this section, we apply our new estimation approach to the Old Faithful Geyser dataset, which consists of times, in minutes, between eruptions of that geyser (found in Yellowstone National Park).  
\tabref{faithful} shows that our estimates (SEM) are close to those obtained by GMM, the method of \cite{hunter2007inference} (SP), the method of \cite{bordes2007stochastic} (SP-EM), and the method of \cite{balabdaoui2014inference} (SLC).
(SEM converged in about a dozen iterations.) 

\begin{table}[h!]
\centering
\begin{tabular}{C{2cm}C{2cm}C{2cm}C{2cm}C{2cm}C{2cm}|}
\toprule
  \multicolumn{1}{C{2cm}}{parameters} &
  \multicolumn{1}{C{2cm}}{GMM} &
  \multicolumn{1}{C{2cm}}{SP} &
  \multicolumn{1}{C{2cm}}{SP-EM} &
  \multicolumn{1}{C{2cm}}{SLC} &
  \multicolumn{1}{C{2cm}}{SEM} \\
\midrule
  \multicolumn{1}{C{2cm}}{$\pi_1$} &
  \multicolumn{1}{C{2cm}}{0.361} &
  \multicolumn{1}{C{2cm}}{0.352} &
  \multicolumn{1}{C{2cm}}{0.359} &
  \multicolumn{1}{C{2cm}}{0.33} &
  \multicolumn{1}{C{2cm}}{0.355} \\

  \multicolumn{1}{C{2cm}}{$\mu_1$} &
  \multicolumn{1}{C{2cm}}{54.61} &
  \multicolumn{1}{C{2cm}}{54.0} &
  \multicolumn{1}{C{2cm}}{54.59} &
  \multicolumn{1}{C{2cm}}{55.5} &
  \multicolumn{1}{C{2cm}}{54.61} \\

  \multicolumn{1}{C{2cm}}{$\mu_2$} &
  \multicolumn{1}{C{2cm}}{80.09} &
  \multicolumn{1}{C{2cm}}{80.0} &
  \multicolumn{1}{C{2cm}}{80.05} &
  \multicolumn{1}{C{2cm}}{80.5} &
  \multicolumn{1}{C{2cm}}{80.5} \\
\bottomrule
\end{tabular}
\caption{\small Parameter estimates for the Old Faithful Geyser dataset.}
\label{tab:faithful}
\end{table}

\bibliographystyle{chicago}
\bibliography{ref}

\end{document}